\documentclass[twocolumn,showpacs,preprintnumbers,amsmath,amssymb,floatfix]{revtex4}
\usepackage{graphicx}
\usepackage{dcolumn}
\usepackage[latin1]{inputenc}
\usepackage{color}

\begin{document}

\title{High Precision Energy Measurements from the Analysis of
Wide Spectral Features. Application to the fluorescence of
YAG:Ce$^{3+}$} 

\author{Miguel Lagos}
\email{mlagos@utalca.cl}
\affiliation{Facultad de Ingenier{\'\i}a, Universidad de Talca,
Campus Los Niches, Camino a los Niches km 1, Curic{\'o}, Chile}

\author{Rodrigo Paredes}
\email{raparede@utalca.cl}
\affiliation{Facultad de Ingenier{\'\i}a, Universidad de Talca,
Campus Los Niches, Camino a los Niches km 1, Curic{\'o}, Chile}

\author{C{\'e}sar Retamal}
\email{ceretamal@utalca.cl}
\affiliation{Facultad de Ingenier{\'\i}a, Universidad de Talca,
Campus Los Niches, Camino a los Niches km 1, Curic{\'o}, Chile}

\date{April 12, 2017}

\begin{abstract}
Advantage is taken of a complete and precise experimental study of
the luminescent properties of yttrium aluminium garnet doped with
Ce$^{3+}$, previously accomplished by other authors, to confirm the
accuracy of the invoked theoretical methods for dealing with the
realistic calculation of the electromagnetic spectra of condensed
phases. The fluorescent spectra at $T=0$ and $T=250\,\text{K}$ of 
YAG:Ce$^{3+}$ were calculated with no adjustable parameter, giving
complete agreement with experiment. The energy released by the
electronic transitions was determined with precision better than
5\% the full width at half maximum of the spectral features.
Thermal quenching of the fluorescent yield is discussed and
calculated in a less accurate way, but anyway showing good
agreement with experiment.
\end{abstract}


\maketitle

Fluorescence is generally associated to transitions of a local
electronic orbital causing a significant structural variation, and
hence a strong perturbation of the dynamics of nuclear degrees of
freedom. The absent, or very weak \cite{Bachmann}, zero-phonon line in
the observed spectra proves in empirical way that few-phonon processes
are infrequent. The wide frequency spectra of the exchanged photons,
the large Stokes shifts and peak asymmetries, and the marked
dependence of the spectral features on the local and long-range
properties of the hosting medium, have all them a common explanation:
the large amount of energy spent in exciting acoustic traveling waves
in the medium hosting the fluorescent molecule
\cite{LagosParedes1,LagosParedes2}.  From a thermodynamic standpoint,
the electronic and electromagnetic radiation fields are two weakly
coupled systems going through a transformation between two well
defined states, in thermal equilibrium with the energy reservoir
constituted by the acoustic vibrational modes of the extended material
medium. By the entropy law, the thermal bath always takes a positive
amount of energy in the average. In photon absorption processes the
radiation field statistically provides the energy delivered to the
thermal bath, and the electron field does in the emission events. Then
the Stokes shift is a direct consequence of the second law of
thermodynamics. Peak asymmetries have a similar general explanation,
since photons carrying an energy excess, or defect, are more probable
in absorption, or emission, processes.

The technology of fluorescence has advanced at an accelerating pace in
a variety of applications \cite{Lakowicz,Mukamel,Lifetech}.  Many of
them follow from the sensitivity of the optical properties of the
fluorophores to the physical attributes of the embedding medium
\cite{Marini,Haidekker,Horng,Diwu,Lai,Safarzadeh,Macgregor}. Despite
these advances, there are still some aspects which deserve deeper
understanding. Recent papers report a theoretical framework able to
reproduce with high numerical precision the temperature dependent
lineshapes \cite{LagosParedes1} and quenching \cite{LagosParedes2} of
the fluorescent spectra of molecules in a condensed environment. The
purpose of this communication is to take advantage of a thorough and
complete study of the optical properties of YAG, previously published
by other authors
\cite{Bachmann}, to briefly show the power of the methods for the
analysis of fluorescent spectra put forward in these papers
\cite{LagosParedes1,LagosParedes2}, and the
convenience of having at hand a good theoretical tool for analyzing
the empirical facts.

Yttrium aluminum garnet ($\text{Y}_3\text{Al}_5\text{O}_{12}$ or YAG)
doped with Ce$^{3+}$ is a phosphor widely applied in white LEDs for
converting blue to yellow light, among many other technical
applications. Fed by the violet and blue lines of the commonly
employed halophosphate phosphor, YAG:Ce$^{3+}$ produces a broad
emission band shifted to yellow which combines with the white light of
the LED phosphor to make it warmer. The experimental study of Bachmann
et al. is particularly useful because exhibits spectra taken at
temperatures from near $T=0$ up to $T=600\,\text{K}$ \cite{Bachmann}.
However, what makes this experiment special is that the resolution was
large enough to clearly observe at 489 nm the zero--phonon line of one
of the two partially superimposed emission bands. This quite unusual
observation in fluorescent spectra gives not only the absolute
energy released in the main electronic transition,
${^2}d\rightarrow{^2F_{5/2}}$, as $E=2.535\,\text{eV}$
\cite{Bachmann,Yanfang}, but also the probability of a zero--phonon
event. A shoulder in the spectral maximum suggests the contribution of
a competing secondary transition, interpreted as
$^2d\rightarrow{^2F_{7/2}}$ \cite{Bachmann,Yanfang}. The relative
intensity of the zero--phonon line at a temperature $T=4\,\text{K}$
is 0.27\% of the total intensity of the main emission band.

In general, the lineshape function is given by the function
\cite{LagosParedes1,LagosParedes2,Lagos1,Lagos2}

\begin{equation}
\begin{aligned}
F(\hbar ck;T)=&\frac{a}{\pi\hbar v_s}
\int_{-\infty}^\infty\, d\tau\, \exp\big\{ -\alpha\big[ J(\tau;T)-
J(\infty ;T)\big]\big\}\\
&\times\exp\bigg\{ i\bigg[\alpha I(\tau )-
\frac{2a}{\hbar v_s}(\hbar ck-E)\tau\bigg]\bigg\}\\
&+\exp\big[-\alpha J(\infty ;T)\big]\,\delta (\hbar ck-E),
\label{E1}
\end{aligned}
\end{equation}

\noindent where $\hbar ck$ is the photon energy, $E$ the energy
difference of the two electronic states involved in the transition,
$a$ is essentially the bond length, and $v_s$ the mean speed of sound
of the acoustic modes of vibration of the medium. The general form of
the auxiliary functions $J(\tau;T)$ and $I(\tau )$ depends on the
symmetry of the surroundings of the optically sensitive orbital
\cite{Lagos2}. For the simplest case of octahedral symmetry they read

\begin{equation}
J(\tau; T)=
\int_0^{aq_D}\,\frac {dx}{x}\, \left( 1-\frac{\sin x}{x}\right)
\coth (\beta x)\sin^2 (\tau x),
\label{E2}
\end{equation}

\begin{equation}
I(\tau )=\frac{1}{2}
\int_0^{aq_D}\,\frac {dx}{x}\,\left( 1-\frac{\sin x}{x}\right)
\sin (2\tau x)\, ,
\label{E3}
\end{equation}

\noindent
with $q_D$ being the cutoff for the wavenumber of acoustic waves, and
the adimensional magnitudes appearing in the lineshape function
$F(\hbar ck;T)$ are given by \cite{LagosParedes1}

\begin{equation}
\alpha =\frac{3(\Delta F)^2}{\pi^2\hbar\rho v_s^3}
\quad\beta =\frac{\hbar v_s}{2 a k_B T}
\quad\tau =\frac{v_s}{2a}t ,
\label{E4}
\end{equation}

\noindent
where $k_B$ is the Boltzmann constant, $T$ the temperature, $\alpha$
an electron--phonon coupling constant, and 

\begin{equation}
J(\infty ; T)=\frac{1}{2}
\int_0^{aq_D}\,\frac {dx}{x}\, \left( 1-\frac{\sin x}{x}\right)
\coth (\beta x).
\label{E5}
\end{equation}

The second term in the right hand side of Eq.~(\ref{E1}) for $F(\hbar
ck;T)$, containing the delta--function is the zero--phonon line, and
the first one the phonon broadened distribution. The lineshape
function $F(\hbar ck;T)$ is normalized as
\cite{LagosParedes1,Lagos1,Lagos2}

\begin{equation}
\int_{-\infty}^{\infty} d(\hbar ck)\, F(\hbar ck;T)=1
\label{E6}
\end{equation}

\noindent
and hence the relative contribution of zero--phonon processes to the
total fluorescent yield is

\begin{equation}
I_{ZPL}=\exp\big[-\alpha J(\infty ;T)\big].
\label{E7}
\end{equation}

Bauchmann et al.~measured $I_{ZPL}=0.0027$ at $T=4\,\text{K}$.  At
$T\approx 0$, Eq.~(\ref{E5}) does not depend on the constant $\beta$
because the hyperbolic function goes rapidly to unity as $\beta$
grows. With practically no precision loss, one can replace $J(\infty
;4\,\text{K})$ by $J(\infty ;0\,\text{K})=0.5817$. This way, the low
temperature data allows to determine the constant $\alpha$

\begin{equation}
0.0027=\exp (-\alpha\times 0.5817)\implies\alpha =10.17.
\label{E8}
\end{equation}

\noindent
It is assumed here that $aq_D=(12\pi^2)^{1/3}$, corresponding to
octahedral symmetry.

On the other hand, the maximum of the main peak of the $T=4\,\text{K}$
spectrum is at $\lambda =523\,\text{nm}$ (2.370 eV).  As the zero
phonon line is at $\lambda = 489\,\text{nm}$ (2.535 eV) the
contribution to the Stokes shift of emission is the difference 0.165
eV. This result is useful to determine the energy scale factor in
Eq.~(\ref{E1}) that puts the maximum of the distribution in the right
place. This gives

\begin{equation}
\frac{\hbar v_s}{2a}=8.8121\times 10^{-22}\,\text{J}
=0.00550\,\text{eV}\, .
\label{E9}
\end{equation}

\noindent
The mean speed of sound in YAG is $v_s=5.47\times 10^3\,\text{m/s}$
\cite{Zhan}. Replacing in Eq.~(\ref{E9}), one obtains that
$a=32.7\,\text{nm}$, which is a very reasonable value. The calculated
distance of substitutional Ce$^{3+}$ in YAG to the nearest oxygen
neighbors is 23--24 nm \cite{Gracia}. The figure for $a$ is closer to
the six Al$^{3+}$ neighbors with coordination quasioctahedral, whose
distances to the Ce$^{3+}$ ion are 30.0 and 33.6 nm \cite{Gracia}.
Hence the assumption of octahedral symmetry for $aq_D$ is consistent
with the predominance of the bonding of the Ce$^{3+}$ ions with their
Al$^{3+}$ next-neighbors.

\begin{figure}[h!]
\begin{center}
\includegraphics[width=8cm]{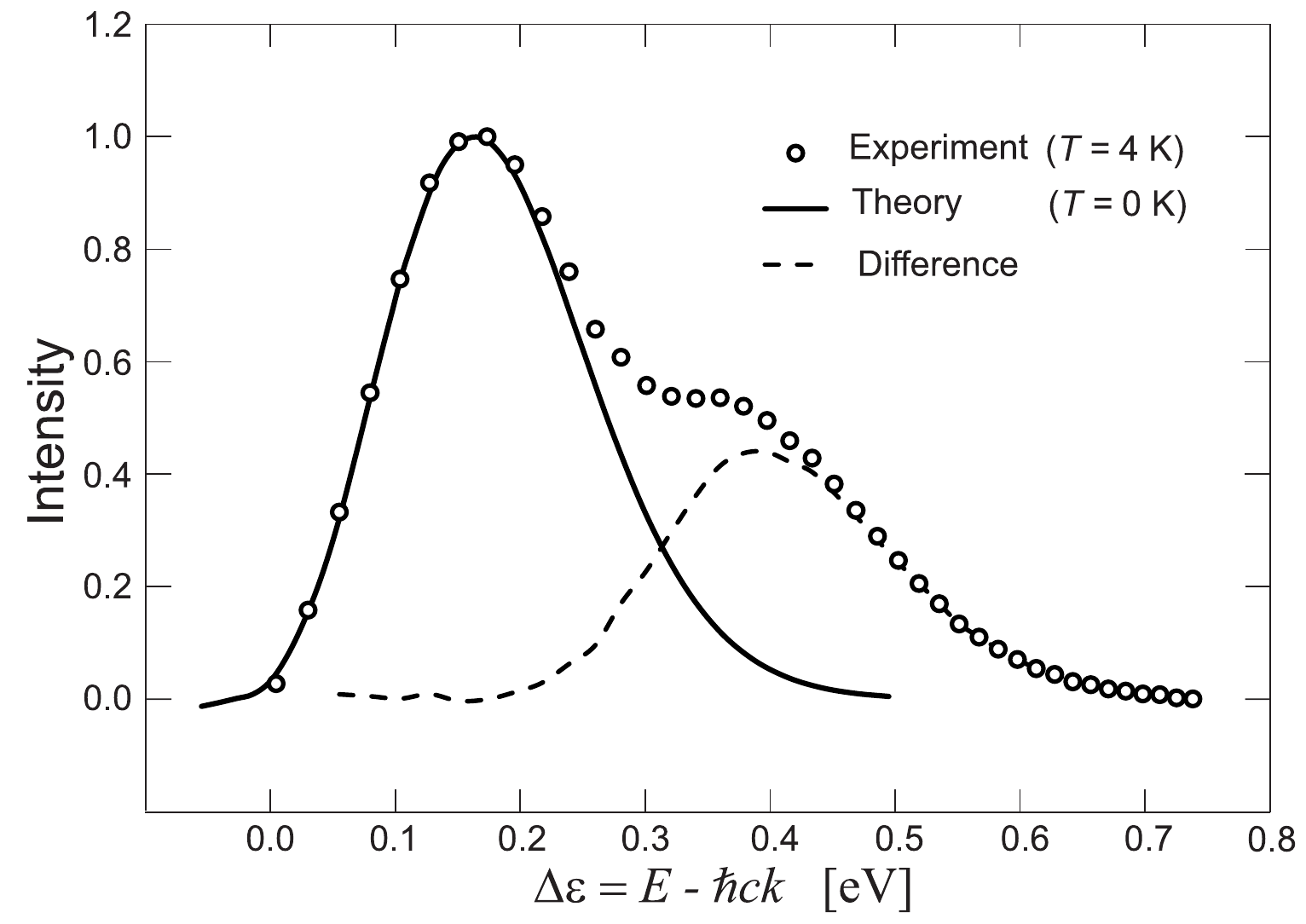}
\caption{\label{Fig1} Open circles represent the fluorescent emission
spectrum of YAG:Ce$^{3+}$ at $T=4\,\text{K}$, as measured by Bauchmann
et al. [1]. The solid line represents Eq.~(\ref{E1}) with constants
determined from the experiment and no fitting parameter.  The origin
of $\Delta\epsilon$ is in the energy of the zero--phonon line of the
transition ${^2}d\rightarrow{^2F_{5/2}}$. The broken line is the
difference, evidencing the existence of a competing transition,
presumably ${^2}d\rightarrow{^2F_{7/2}}$.}
\end{center}
\end{figure}

Knowing the two constants $\alpha$ and $\hbar v_s/(2a)$ the lineshape
function (\ref{E1}) can be evaluated for $T=0$. The resulting curve is
shown in Fig.~\ref{Fig1}, together with the experimental data of
Bauchmann et al.~for $T=4\,\text{K}$. The coincidence between theory
and the main feature of the experimental spectrum (attributed to the
$5d^1\rightarrow {^2F}_{5/2}$ transition \cite{Yanfang}) is
remarkable, particularly because the theoretical curve has no
adjustable parameter.

\begin{figure}[h!]
\begin{center}
\includegraphics[width=8cm]{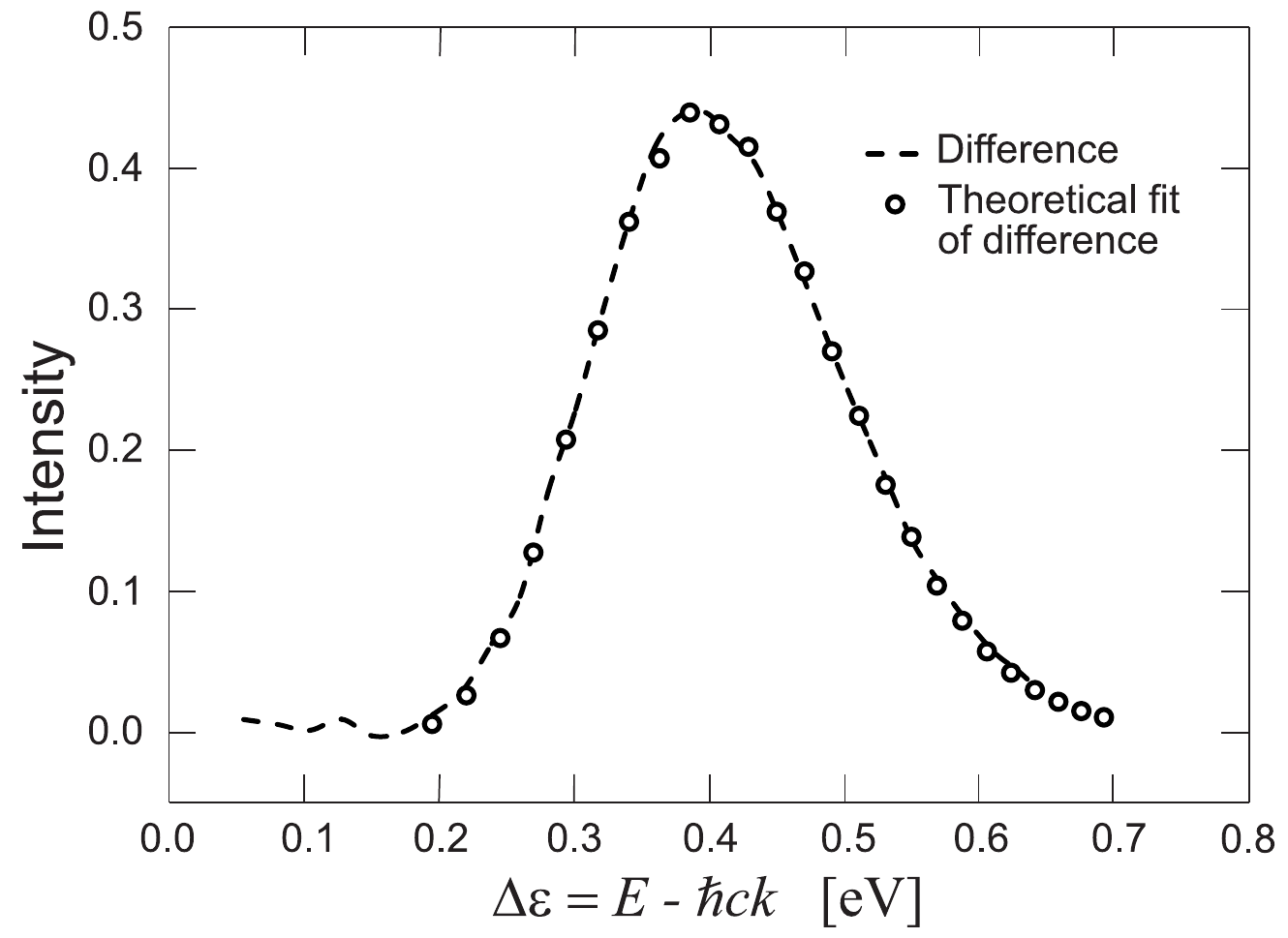}
\caption{\label{Fig2} The broken line depicts the difference between
the theoretical curve and experimental data shown in Figure 1. Open
circles represent Eq.~(\ref{E1}) with the coupling constant $\alpha$
as a fitting parameter and $\Delta\epsilon$ displaced in 0.192 eV in
order to describe the second transition
${^2}d\rightarrow{^2F_{7/2}}$.}
\end{center}
\end{figure}

The broken lines in Figs.~\ref{Fig1} and \ref{Fig2} represent the
difference between the experimental data and the theoretical curve for
$\alpha =10.17$ and $\hbar v_s/(2a)= 0.00550\,\text{eV}$. Interpreting
this new data set as the emission attributed to the transition
$5d^1\rightarrow {^2F}_{7/2}$ \cite{Yanfang}, a theoretical fit of it
is essayed with the lineshape function (\ref{E1}), with the coupling
constant $\alpha =12.3$, same energy scale constant $\hbar
v_s/(2a)=0.00550\,\text{eV}$, and the photon energies shifted in
$0.192\,\text{eV}$. The new zero phonon line is then at
$E'=2.343\,\text{eV}$, which is the net energy released in the second
transition. As shown by Fig.~\ref{Fig2}, the correspondence between
theory and experiment is again very good. The displacement of the
energy transfers affects also the zero--phonon line and the net energy
released in the second transition. Table 1 summarizes the
situation.

\begin{table}[h!]
\begin{center}
\renewcommand{\arraystretch}{1.5}
\setlength{\tabcolsep}{7pt}
\caption{\label{table1} Energy released in the electronic transitions.}
\begin{tabular}{|c|c|} \hline
Transition & Energy [eV] \\ \hline
${^2}d\rightarrow{^2F_{5/2}}$ &  2.535 \\ \hline
${^2}d\rightarrow{^2F_{7/2}}$ &  2.343 \\ \hline
\end{tabular}
\end{center}
\end{table}

The empirical data we are comparing with are normalized by setting the
maxima of the distributions at unity. The theoretical curve of
Fig.~\ref{Fig1} was normalized accordingly, dropping the factor
preceding the integral in Eq.~(\ref{E1}) and multiplying it by a
weight factor 12.00 to put the maximum at unity. The fit of the
theoretical curve and the data in Fig.~\ref{Fig2} was attained by the
same procedure with a weigth factor 5.71. Thus the ratio of the
intensities of the two emission bands is close to 2:1.

The constants $\alpha$ and $\hbar v_s/(2a)$ are enough to determine
the line shape at any temperature through Eq.~(\ref{E1}). The
integrated intensities, or quantum yields, constitute a different
matter because competitive non radiative decay channels may exist,
giving rise to temperature dependent quenching of the quantum
efficiency. The shape of the absorption and emission bands, given by
Eq.~(\ref{E1}), are not affected because quenching is in general due
to the simultaneous draining of the excited states by an alternate non
radiative process \cite{LagosParedes2}. The emission probabilities per
unit time of photons of different energies are then affected the same
way, and their relative values are not changed.

\begin{figure}[h!]
\begin{center}
\includegraphics[width=8cm]{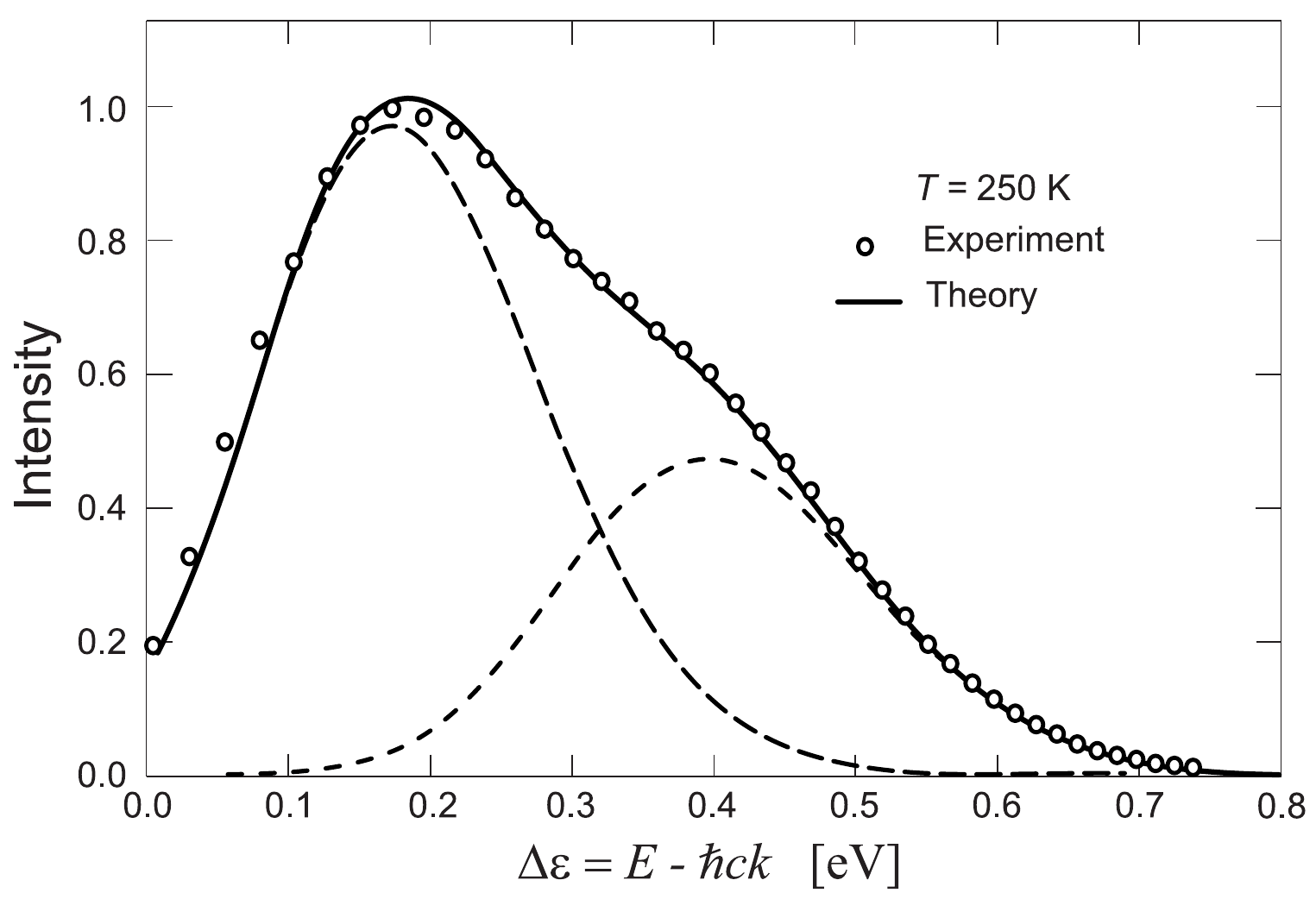}
\caption{\label{Fig3} Emission intensity at $T=250\,\text{K}$. Broken
lines depict the contribution of the two transitions, as given by
Eq.~(\ref{E1}) with the same constants determined from the analysis
of the $T=0$ data. Circles represent the data of Bachmann et al.
\cite{Bachmann}}
\end{center}
\end{figure}

The calculations for $T=0$ were repeated for $T=250\,\text{K}$,
yielding the curves of Fig.~\ref{Fig3}. They were obtained conserving
$\alpha = 10.17$ and 12.3 for the 2.535 eV and 2.343 eV transitions,
respectively, and $\hbar v_s/(2a)=0.0550\,\text{eV}$. This gives
$\beta =0.2553$ for $T=250\,\text{K}$. The broken lines represent the
contributions of the two transitions, and the solid line is the sum of
both with weighting factors 13.95 and 7.50. Although the ratio of the
intensities varied a little with respect to the case $T=0$, it is
still close to 2:1. Actually, the ratio between the intensities of the
two contributions differs from 2:1 by 5\% and -7\% at $T=0$ and
$T=250\,\text{K}$, respectively. This suggests that the transition
contributing the most may be connected with the four coplanar
Al$^{3+}$ next-neighbors to the Ce$^{3+}$ ion, and the remaining two
Al$^{3+}$ ions of the quasi--octahedral array be connected with the
weaker contribution.

\begin{figure}[h!]
\begin{center}
\includegraphics[width=8cm]{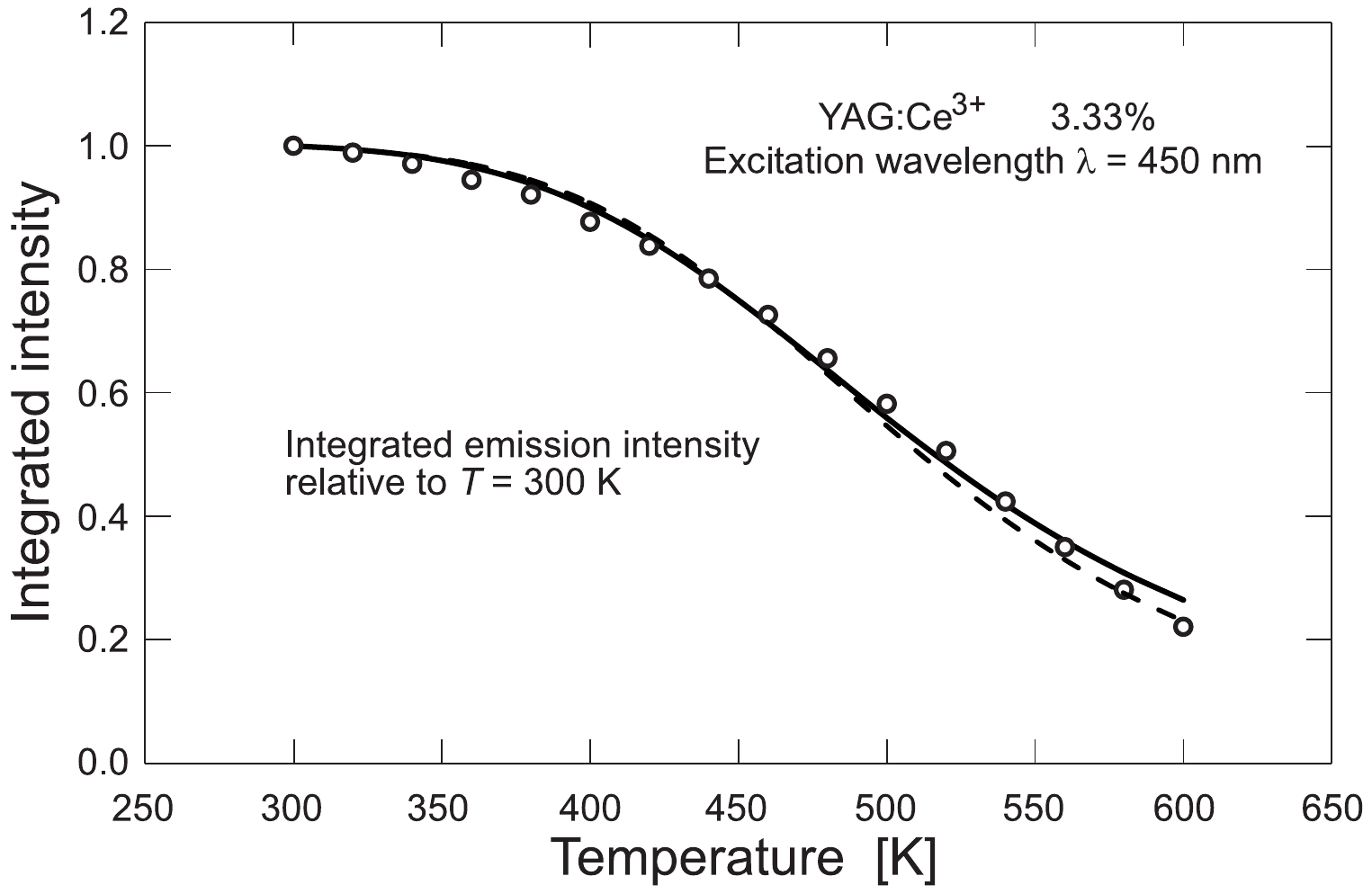}
\caption{\label{Fig4} Temperature dependence of the quantum yield of
the fluorescence of 3.33\% concentration YAG:Ce$^{3+}$, excited by
light of 450 nm wavelength \cite{Bachmann}. Circles stand for
experiment and the solid line represents the approximate
Eq.~(\ref{E10}) with $p=2.20\times 10^{-7}\,{\text{J}}^{1/2}$ and
$\varepsilon =0.350\,\text{eV}$. The meaning of the broken line is
discussed in the text. Experimental and theoretical results show some
discrepancies because Eq.~(\ref{E10}) is an asymptotic expression
valid under conditions which are not fully satisfied here.}
\end{center}
\end{figure}

A deep study on the problem of quenching and its temperature
dependence has been recently published \cite{LagosParedes2}. The
theoretical investigation of the thermal quenching of YAG:Ce$^{3+}$ is
particularly interesting because has been the subject of thorough
experimental study of Bachmann et al. The predicted quantum yield
given by Eq.~(\ref{E1}) is always unity, as expressed by
Eq.~(\ref{E6}), independent of the temperature and particularities of
the hosting medium. This is because the derivation of Eq.~(\ref{E1})
assumes a single decay channel and drops terms of the system
Hamiltonian which becomes significant when the coupling of the
electronic and configurational variables become strong enough, as
occurs in fluorescent solids or molecules in a condensed medium. These
terms do not contain variables of the radiation field and give rise to
non radiative decay channels which compete with the radiative
contribution \cite{LagosParedes2}.

The temperature dependent integrated intensity, or quantum yield, of
YAG:Ce$^{3+}$ fluorescence deserves a detailed study. By the time
being we will simply apply the asymptotic equation for the ratio of
the quantum yield $\phi (T)$ at two temperatures, $T$ and a reference
temperature $T_0$ \cite{LagosParedes2},

\begin{equation}
\frac{\phi (T)}{\phi (T_0)}=
\frac{1+\dfrac{p}{\sqrt{k_BT_0}}
\exp\bigg(-\dfrac{\varepsilon}{k_BT_0}\bigg)}
{1+\dfrac{p}{\sqrt{k_BT}}
\exp\bigg(-\dfrac{\varepsilon}{k_BT}\bigg)},
\label{E10}
\end{equation}

\noindent
which holds for high temperatures and strong enough electron--phonon
coupling. The latter is made apparent by the breadth of the spectral
maximum. Eq.~(\ref{E10}) has proven to reproduce with great accuracy
the temperature dependent yield efficiency of dissolved
8-methoxypsoralen, which displays a single fluorescent maximum of 0.6
eV full width at half maximum (FWHM) \cite{LagosParedes2}.  However,
the maxima exhibited by YAG:Ce$^{3+}$ do not go beyond 0.25 eV FWHM,
and then the accuracy of the simple Eq.~(\ref{E10}) in the present
case may be less than perfect. Fig.~\ref{Fig4} shows the data of
Bauchmann et al.  for the ratio expressed in the right hand side of
Eq.~(\ref{E10}) with $T_0=300\,\text{K}$ (circles). The solid line
represents Eq.~(\ref{E10}) with the constants $p=2.20\times 10^{-7}\,
\text{J}^{1/2}$ and $\varepsilon =5.61\times 10^{-20}\,
\text{J}=0.350\,\text{eV}$. The fit of the theoretical curve to the
experimental data is fairly good for values of the constants which
have the expected magnitudes. However small but systematic deviations
do exist. Beside the rather insufficient width of the spectral
maximum, the temperature range is another source of errors. The mean
speed of sound in YAG is $v_s=5.47\times10^{3}\,\text{m/s}$
\cite{Yanfang}, which is a rather high value. Hence the Debye
temperature is expected to be high as well, higher than 300 K. As
the Debye temperature is the reference for temperatures,
Eq.~(\ref{E10}) is expected to be not very precise in the interval
$300\,\text{K}<T<600\,\text{K}$. The temperature dependent quantum
yield curves reported by Bachmann et al. vary with the Ce$^{3+}$
concentration and wavelength of the exciting light.  Independence of
these variables is attained for just the highest concentration
(3.33\%). The effect is presumably due to the formation of small
clusters of the substitutional Ce impurities. The local distortions
they produce in the lattice give rise to an attractive
interaction. These hypothetical clusters would be less regular for
lower concentrations, giving rise to the dependence on exciting
wavelength and concentration. By this reason Fig.~\ref{Fig4} takes the
data of the higher 3.33\% concentration.

The broken line in Fig.~\ref{Fig4} represents the curve given by an
equation similar to Eq.~(\ref{E10}), but with the factor
$p/\sqrt{k_BT}$ replaced by the numerical constant 3420 and
$\varepsilon =5.75\times 10^{-20}\,\text{J}=0.359\,\text{eV}$. This
was done because this kind of equation has been essayed in the
literature for fitting the temperature dependence of the thermal
quenching of luminescence. Also, the broken line in Figure 4 shows
that, although multiplying a strongly varying exponential factor, the
temperature dependence of the pre--exponential factor is not
completely lost and can be elucidated from the analysis of the data.

\vskip 16pt
\noindent
{\bf Acknowledgments.} This work was partially supported by the
research fund Enlace--Fondecyt, Direcci{\'o}n de Investigaci{\'o}n,
Universidad de Talca, Chile (M. L. and R. P.), also by CONICYT,
FONDECYT , Chile, under Grant No. 1131044, (R. P.). C. R. acknowledges
financial support from CONICYT, FONDECYT Iniciaci{\'o}n, Chile, under
Grant No. 11130624.

\end{document}